# Ion cyclotron emission from fusion-born ions in large tokamak plasmas: a brief review from JET and TFTR to ITER

R O Dendy[1,2] and K G McClements[1]

[1]*CCFE, Culham Science Centre, Abingdon, Oxfordshire OX14 3DB, U.K.*

[2]*Centre for Fusion, Space and Astrophysics, Department of Physics,
Warwick University, Coventry CV4 7AL, U.K.*

**Abstract**

Ion cyclotron emission (ICE) was the first collective radiative instability, driven by confined fusion-born ions, observed from deuterium-tritium plasmas in JET and TFTR. ICE comprises strongly suprathermal emission, which has spectral peaks at multiple ion cyclotron harmonic frequencies as evaluated at the outer mid-plane edge of tokamak plasmas. The measured intensity of ICE spectral peaks scaled linearly with measured fusion reactivity in JET. In other large tokamak plasmas, ICE is currently used as an indicator of fast ions physics. The excitation mechanism for ICE is the magnetoacoustic cyclotron instability (MCI); in the case of JET and TFTR, the MCI is driven by a set of centrally born fusion products, lying just inside the trapped-passing boundary in velocity space, whose drift orbits make large radial excursions to the outer mid-plane edge. Diagnostic exploitation of ICE in future experiments therefore rests in part on deep understanding of the MCI, and recent advances in computational plasma physics have led to substantial recent progress, reviewed here. Particle-in-cell simulations of the MCI, with fully kinetic ions and electrons, were reported in 2013, using plasma parameters for JET ICE observations. The hybrid approximation for plasma simulations, where ions are treated as particles and electrons as a neutralising massless fluid, was then applied and reported in 2014. These simulations extend previous studies deep into the nonlinear regime of the MCI, and corroborate predictions by linear analytical theory, thereby strengthening further the link to ICE measurements. ICE is a potential diagnostic for confined alpha-particles in ITER, where measurements of ICE could yield information on energetic ion behaviour supplementing that obtainable from other diagnostics. In addition, it may be possible to use ICE to study fast ion redistribution and loss due to MHD activity in ITER.

## 1. A brief history of ion cyclotron emission from fusion plasmas

Understanding the physics of populations of energetic ions, born in fusion reactions between thermal ions, is central to the future exploitation of magnetically confined plasmas for energy generation. Ion cyclotron emission (ICE)[1,2] was the first collective radiative instability, driven by confined fusion-born ions, that was observed from deuterium-tritium plasmas[3-6] in both JET and TFTR. Intensely suprathermal emission, strongly peaked at the frequencies of sequential ion cyclotron harmonics evaluated at the outer mid-plane edge, was initially detected using heating antennas in receiver mode on JET and using probes in TFTR. The measured intensity of ICE spectral peaks scaled linearly with measured fusion reactivity: both between different plasmas spanning a factor of a million in fusion reactivity[3], and during single plasma pulses following the rise and fall of fusion reactivity over time[3,5]. Soon after these observations were reported, linear analytical theory together with particle orbit calculations suggested[7-11] that the emission mechanism is the magnetoacoustic cyclotron instability (MCI). The MCI was first identified theoretically by Belikov and Kolesnichenko[12] before it was observed in JET and TFTR tokamak plasmas, where it is driven by a subset of centrally born fusion products that lie just inside the trapped-passing boundary in velocity space, whose existence was anticipated by Stringer[13]. In both JET and TFTR, the drift orbits of these ions make large radial excursions[3,4] to the outer mid-plane edge. Since the local fusion birth rate is very small at this location, the predominant local energetic particle population comprises the super-Alfvénic centrally born ions that pass through on their drift orbits: hence there is a local population inversion in velocity space, and consequently scope for fast collective relaxation through the MCI. More recently, in other large tokamak plasmas, ICE has been used as a diagnostic for lost fast ions due to MHD activity in DIII-D[14,15], and has been studied in detail in ASDEX-Upgrade[16,17] and JT-60U[18,19]. In the latter two tokamaks, ICE is seen at edge cyclotron harmonics of ion species that include the products of fusion reactions in pure deuterium, namely protons, tritons, and helium-3 ions. This is the initial phenomenology reported in Ref.[1], but with diagnostic power and aspects of plasma performance greatly improved over the intervening quarter-century. ICE from sub-Alfvénic beam ions, injected for plasma heating purposes at tens of keV, has also been observed from tokamak plasmas, and the corresponding variant of the MCI is understood analytically[20]. ICE is also used as a diagnostic of lost fast ions in large stellarator-heliotron plasmas in LHD[21].

ICE is of further interest as a potential diagnostic for the evolving fusion-born alpha-particle population in future deuterium-tritium plasmas in JET and ITER[11]. In addition, it may be possible to use ICE to study fast ion redistribution and loss due to MHD activity in ITER. Among other examples, clear evidence has been found of links between ICE and sawteeth[2] and ELMs[3] in JET, fishbones[15] in DIII-D, and TAEs[21] in LHD. Deep understanding of the physics of the MCI is therefore of great practical interest for magnetic confinement fusion and, as reviewed here, advances in computational plasma physics have led to substantial recent progress. The particle-in-cell (PIC) code described in Ref.[22] was used to perform simulations[23] of the MCI which carried the instability into its nonlinear regime for the first time. These PIC simulations modelled fully kinetic ions and electrons, together with the self-consistent electric and magnetic fields, for plasma parameters aligned to relevant JET conditions. The results[23] strengthened the link to ICE observations, and this was reinforced by subsequent simulations of the MCI over even longer physical time durations[24] using a code implementing the hybrid approximation[25], where ions are treated as particles and electrons as a neutralising massless fluid. These simulations[23,24] corroborate predictions from linear analytical theory where applicable and, by extending previous studies deep into the nonlinear regime of the MCI[24], confirm that the MCI underlies the ICE observations from energetic ions in large magnetically confined plasmas.

Electromagnetic radiation with properties similar to ICE has been detected in solar-terrestrial plasmas at two distinct locales, under conditions where it is likely to arise from the MCI[26-28], driven by observed local populations of minority energy ions. In astrophysics, the possibility of related effects at supernova remnant (SNR) shocks was originally noted in Ref.[29]. Here the minority energetic ion population comprises cosmic rays undergoing pre-acceleration from thermal to mildly relativistic energies. Recent PIC simulations show that, in the region of turbulent plasma immediately downstream of SNR shocks, the spectral distribution of wave energy is concentrated at sequential ion cyclotron harmonic peaks[30]. These simulations describe plasmas containing two ion species, protons and alpha-particles, and the full range of relative concentrations is examined. The driving population of energetic ions comprises particles from upstream that have passed through the collisionless shock into the downstream plasma.

Finally, we note that study of the MCI addresses a fundamental question in classical electrodynamics: if there is a minority drifting ring-beam population of energetic ions in a plasma, then how do this population, the plasma, and the self-consistently excited fields evolve and interact over the long term?

## 2. Modelling ion cyclotron emission

The present state-of-the-art of ICE interpretation is encapsulated in Fig.1, which is reproduced from Ref.[24]. This combines three elements: observational ICE spectra from a JET deuterium-tritium plasma; calculations of linear analytical growth rates for the MCI using corresponding parameter values; and the results of self-consistent nonlinear kinetic simulation using a hybrid model and exploiting contemporary high performance computation. The hybrid model treats ions as kinetic particles, whose positions evolve continuously in physical and velocity space, each acted upon by the local Lorentz force. Electron physics is modelled in terms of a charge-neutralising fluid. The magnetic field is updated using Faraday's law, and the electric field is updated using the electron fluid momentum equation in the limit of zero electron inertia[24,25]. Like PIC models[22,23], hybrid models fully resolve ion kinetics, and thus include (unlike gyrokinetic models) the full orbit gyromotion of the ions. This enables the study of instabilities that evolve fast, on timescales which are rapid compared to those of gyro-averaged or fluid phenomena, and are very rapid compared to collisional slowing-down. In the simulations reviewed here, the majority thermal plasma is supplemented by an energetic minority alpha-particle population, whose concentration relative to the thermal ions is typically in the range = $10^{-3}$ to $10^{-5}$. This population is represented in velocity space by a ring-beam with characteristic energy 3.5MeV, which is the birth energy of alpha-particles in deuterium-tritium plasma. The simulation domain is three-dimensional in velocity space and one dimensional in real space (1D3V). The spatial domain can be set at arbitrary orientation with respect to the direction of the background magnetic field.

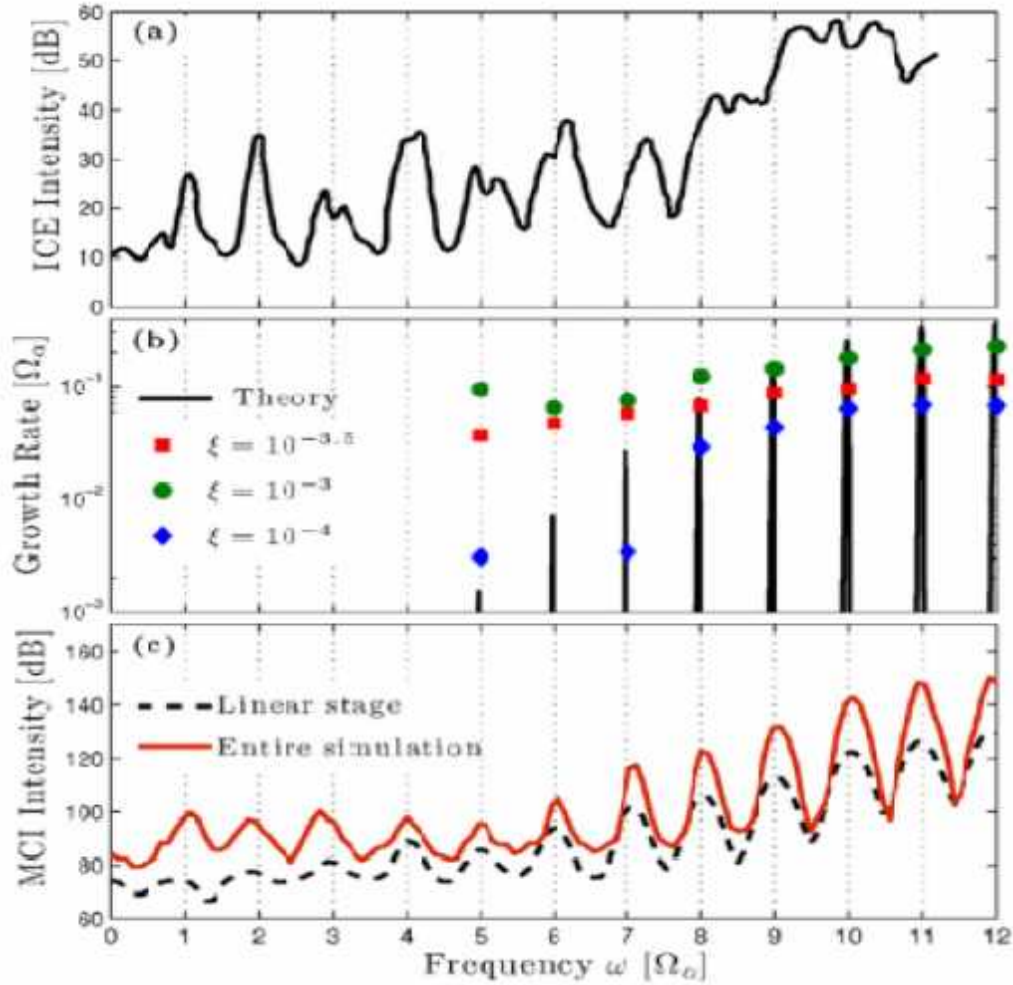

Fig.1, reproduced with permission from Ref.[24]. Experimental ICE spectra compared with linear analytical growth rates for the MCI and with computed wave intensities obtained from fully nonlinear MCI simulations using the hybrid model. Frequencies and growth rates are in units of the alpha-particle cyclotron frequency, $\Omega$. Top panel: Measured ICE intensity from JET deuterium-tritium plasma 26148. Middle panel: Analytical linear growth rate for the MCI for the number density ratio = $10^{-3}$ (black line); together with the early time (corresponding to linear) growth rate of the MCI inferred from the hybrid simulations for the three values, = $10^{-3}$, $10^{-3.5}$, $10^{-4}$ (coloured markers). Bottom panel: Intensity of the electromagnetic field component $B_z$ in the hybrid simulation with number density ratio = $10^{-3}$, at earlier (dashed black line) and later (red line) times.

The hybrid simulations[24] of the MCI and ICE were preceded chronologically by 1D3V PIC simulations[23], in which electrons are treated fully kinetically as well as ions, and the correct physical mass ratio for protons and electrons $m_p/m_e = 1836$ was used. The PIC approach has the advantage of maximum physical fidelity, and a corresponding disadvantage in that it is computationally expensive, such that the evolution of the MCI can be followed only for about five gyro-periods of the alpha-particles. It is found that the MCI unfolds so fast that this PIC simulation duration is sufficient to take one through the linear phase of the instability, but not far into the nonlinear phase of the MCI, which is more fully explored in the hybrid simulations. The fluid treatment of the electrons in the hybrid model also maps to the analytical treatment[5-12,20] of the MCI, which derives from multi-species dielectric tensor elements whose electron component is calculated in the cold plasma approximation with the addition of a Landau resonance term.

A consistent picture emerges from Fig.1: agreement between ICE observations and hybrid simulation is good, especially once the hybrid simulation reaches the nonlinear regime, as seen in the bottom panel; and there is also an appropriate mapping to analytical theory of the MCI. The fact that the hybrid model is able to follow the MCI deeper into its nonlinear phase, enables this treatment[24] to capture additional aspects of the observed ICE signal. Comparison of the two traces for the field intensity in the hybrid simulations in Fig.1(bottom) with the measured ICE signal in Fig.1(top) shows that only the solid red trace, which encompasses the nonlinear phase, captures clearly the lowest three observed cyclotron harmonic peaks. The dashed black line, obtained at earlier times in the hybrid simulation, aligns with the corresponding trace for the PIC simulations, shown in the bottom panel of Fig.1 of Ref.[23]. These simulations, namely early time – hence linear phase – hybrid and PIC, show relatively weak excited field amplitudes at the lowest cyclotron harmonics. In this, they align also with analytical theory which predicts linear stability for the lowest harmonics for these plasma parameter values, as seen in the middle panel of Fig.1.

Recent analysis[24] of the outputs of the hybrid code shows that the drive responsible for low cyclotron harmonics arises initially from nonlinear coupling between neighbouring modes of the electromagnetic field at higher cyclotron harmonics, that are excited in the early phase. This is shown in Fig.2, which shows bicoherence plots at early and late times. Bispectral analysis of Fourier transformed turbulent fields is an elegant method[31-34] for quantifying the degree of nonlinear wave-wave coupling present; that is, the class of interaction that occurs in weak turbulence, where approximate normal modes can be identified and interact. The left panel of Fig.2, taken at a time towards the end of the linear phase of the MCI in the hybrid simulation, shows the strongest nonlinear coupling to be between neighbouring cyclotron harmonics in the range eight to twelve. Nonlinear beating between these waves drives up the lowest harmonics, which are also seen to be coupled to higher harmonics. This is a classic signature of nonlinear three-wave coupling, which proceeds into the right panel of Fig.2, taken at a time in the nonlinear phase of the MCI. The coupling strength between low and high harmonics has now become comparable to that between neighbouring high harmonics.

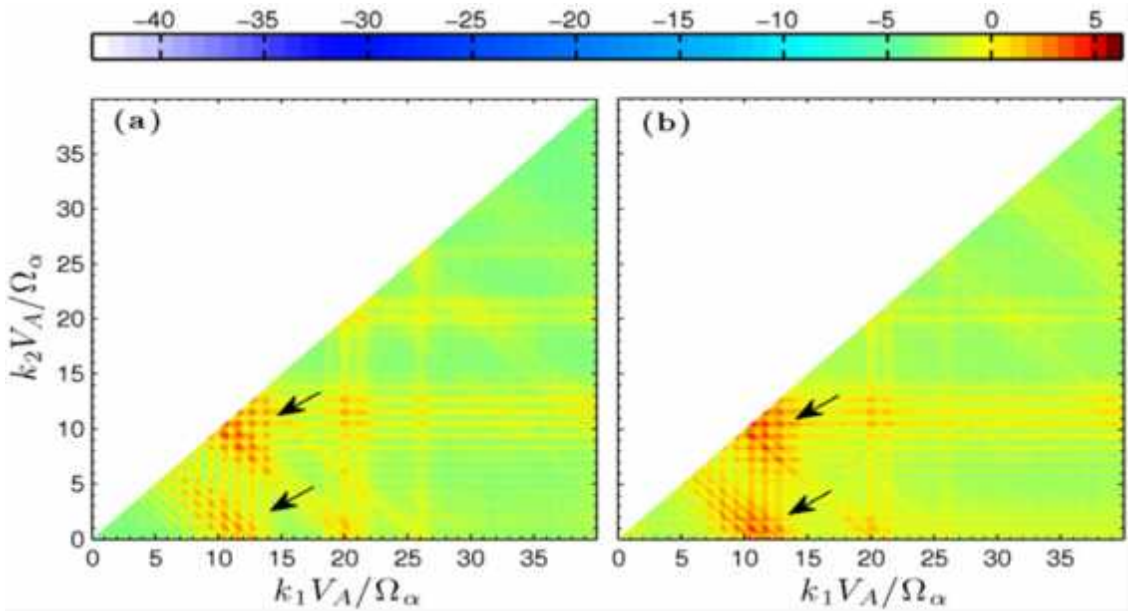

Fig.2, reproduced with permission from Ref.[24]. Origin of strong spectral peaks at the lowest cyclotron harmonics. The non-normalized self-bicoherence of the $B_z$ component is plotted in two-dimensional wavenumber space using a $\log_{10}$ colour scale. The red end of the colour scale indicates significant coupling between different modes at $k_1$ and $k_2$. Left and right panels show the non-normalised self-bicoherence of $B_z$ in the linear and nonlinear stages of the MCI, respectively. Since the excited waves satisfy approximately the fast wave dispersion relation $\check{S} = kV_A$ where $V_A$ is the Alfvén speed, $k_1V_A/\Omega$ is approximately equal to the cyclotron harmonic number. The arrows in both panels point to the locations in the ($k_1$, $k_2$) plane at which coupling is strongest between the $k_1$ and $k_2$ modes.

The foregoing illustrates how recent nonlinear kinetic simulations of plasma that contains an energetic minority ion population and is undergoing the MCI have augmented the original linear analytical treatment, and further strengthened the observational link to ICE. Fourier transforming the electric and magnetic fields that are excited in these kinetic simulations has provided further reinforcement, by demonstrating that their dispersive properties[23,24] are indeed those of the modes assumed in analytical theory, namely fast Alfvén and cyclotron harmonic waves.

## 3. Conclusions

This review has provided a brief account of why it is thought probable that the plasma physics process underlying observations of ion cyclotron emission (ICE) from large tokamak plasmas is the magnetoacoustic cyclotron instability (MCI). This was strongly suggested by the original linear analytical theory approach to ICE from deuterium-tritium plasmas in JET and TFTR[1-7]. It is confirmed by the recent large scale kinetic simulations using PIC and hybrid codes[14,15] reviewed here. The 1D3V simulation domain employed does not include toroidal geometry and the associated compressional Alfvén eigenmode structures[35,36]. Nevertheless the 1D3V nonlinear kinetic level of description appears sufficient for capturing most of the key observed phenomenology. This depth of understanding of the emission mechanism is essential if ICE is to be exploited as a diagnostic of confined and lost fusion alpha-particles in ITER, as has been proposed[37]. The simulations indicate that most of the key physics of the MCI, which in turn determines the observational characteristics of ICE, unfolds on the rapid timescale of a few ion cyclotron periods, which is extremely short compared to the evolution timescale of the overall alpha-particle population in quasi-steady state. The rapidity of the fully nonlinear MCI also accounts for the unreasonable success of linear analytical theory, which leaves its imprint on ICE observations. As noted in Section 1, these results have applications beyond fusion research, notably in space and astrophysical plasma physics.

Looking to the future, we have seen that ion cyclotron emission (ICE) is ubiquitous from energetic ion populations, especially fusion-born ions, in the largest tokamak plasmas – JET, TFTR, JT-60U, ASDEX-U and DIII-D – and in the largest stellarator-heliotron plasma LHD. ICE arises spontaneously and is detected using passive diagnostic equipment. The physics underlying the excitation process for ICE, namely the magnetoacoustic cyclotron instability (MCI), is understood with considerable analytical and computational fidelity. It can therefore be argued[37] that any future fusion experiments involving deuterium-tritium plasmas, notably in JET and in due course ITER, would benefit from a modest effort to detect ICE. This would offer a unique channel to understanding confined alpha-particle physics, provide continuity with the knowledge base from past deuterium-tritium experiments, and sustain an interesting interface with wider science.

## Acknowledgments

It is a particular pleasure to thank: Geoff Cottrell (JET/Culham) for drawing the first JET fusion-product-driven ICE spectrum to the first author's attention in 1986, and for much subsequent collaboration; Dick Majeski (Princeton) for experimental and interpretive collaboration during the deuterium-tritium campaigns at TFTR in the 1990s; and Sandra Chapman, together with our PhD students James Cook and Leopoldo Carbajal (all Warwick), for collaborations exploiting HPC for ICE interpretation from 2010 onwards. This work was part-funded by the RCUK Energy Programme [under grant EP/I501045] and the European Communities. To obtain further information on the data and models underlying this paper please contact PublicationsManager@ccfe.ac.uk. The views and opinions expressed herein do not necessarily reflect those of the European Commission.